\documentclass[a4paper]{amsart}
\usepackage{amssymb,pstricks,graphicx,subfigure,mathrsfs,hyperref}

\newtheorem{definition}{Definition}

\newcommand{\diag}{\mathrm{diag}}

\begin{document}
\title{Hypergeometric solutions to ultradiscrete Painlev\'e equations}
\author{Chris Ormerod}
\begin{abstract}
We propose new solutions to ultradiscrete Painlev\'e equations that cannot be derived using the ultradiscretization method. In particular, we show the third ultradiscrete Painelev\'e equation possesses hypergeometric solutions. We show this by considering a lift of these equations to a non-archimedean valuation field in which we may relax the subtraction free framework of previous explorations of the area. Using several methods, we derive a family of hypergeometric solutions.
\end{abstract}
\maketitle
The area of integrable discrete mappings has blossomed within the last few years. Integrable discrete versions of the Painlev\'e equations is an area of active research\cite{SC1}. Analogous to the continuous Painlev\'e equations, as a hallmark of integrability, the discrete Painlev\'e equations admit rational solutions \cite{Kaji1} and also hypergeometric solutions \cite{Kaji2}. The ultradiscrete Painelev\'e equations are obtained via a limiting process called ultradiscretization \cite{ultra}. The ultradiscretization process sends a rational function of some variables $f(a_1,\ldots, a_n)$ to a rational function over a semiring in a new set of ultradiscrete variables by the following limit
\[
F(A_1,\ldots, A_n) = \lim_{\epsilon \to 0} \epsilon \log(f(a_1, \ldots, a_n)
\]
where the new ultradiscrete variables are related to the old variables by the equations $a_i = e^{A_i/\epsilon}$. Such a process successfully related an integrable cellular automata known as the box-ball system \cite{boxball} to integrable $q$-difference equations \cite{ultra}. Roughly speaking it is a transformation bringing the following binary operations to their ultradiscrete equivalent given by
\begin{subequations}\label{corr}
\begin{eqnarray}
a + b &\to& \max(A,B)\\
ab &\to& A+ B\\
a/b & \to & A - B 
\end{eqnarray}
\end{subequations}
where there is no analog of subtraction. The inequivalence of subtraction often restricts any calculations to a subtraction free framework, in that many of the methods for studying the resultant equations come from subtraction free methods associated with the mapping it is derived from.

One derives an ultradiscrete Painlev\'e equation by applying the ultradiscretization method to a subtraction free version of the $q$-difference analogs of a Painlev\'e equation\cite{Gramani_1997}.There are ultradiscrete analogs of all six of the Painlev\'e equations \cite{ultimate}. We will focus on the ultradiscretization of the equation
\begin{equation}\label{qP3}
  w(q t) w(t/q) = \frac{a_3a_4(w(t)+a_1t)(w(t)+a_2t)}{(w(t)+a_3)(w(t)+a_4)}
\end{equation}
which is given by
\begin{eqnarray}
\label{udP3}\overline{W} +\underline{W} &=& A_3 + A_4 + \max(W, A_1 + T)+ \max(W,A_2 + T) \\
&& - \max(W,A_3) - \max(W,A_4) \nonumber.
\end{eqnarray}
These equations are known as $q$-$\mathrm{P}_{\mathrm{III}}$ and $u$-$\mathrm{P}_{\mathrm{III}}$ respectively. There is evidence that such equations are integrable, such as the existence of a Lax Pair \cite{Corm1}. There are known rational solutions to such equations \cite{Takahashi_1997}. These rational solutions come as ultradiscretized subtraction free rational solutions to the $q$-difference equations. There is also an attempt to relax the subtraction free nature by considering alternate ultradiscretizations \cite{sinh}, yet this method does introduce further difficulties and are considered in a different class to that of the ultradiscrete Painlev\'e equations. What is not known is the existence of the hypergeometric solutions of the ultradiscrete Painlev\'e equations.

Using a construction of a field $\Omega$ with a non archimedean valuation $\nu$, we lift the Painelv\'e equations from the max-plus semiring, $S$ , to equations over $\Omega$ in which $\nu$ acts as a homomorphism of subsemiring of $\Omega$. Under some set of conditions beyond the subtraction free nature of a function, the mapping $\nu$ is a homomorphism. We show an application of this by the derivation of the hypergeometric solutions of \eqref{udP3} which we regard as a derivation of a solution which is not derived using the ultradiscretization method.

In section \ref{sec:tools}, we introduce the max-plus semiring and review some key concepts such as valuation rings. We also construct the main tool, which is known as the inversible max-plus algebra, $\Omega$. In section \ref{sec:hyper} we review some methods of obtaining the $q$-hypergeometric functions for \eqref{qP3}. These methods include an the introduction to the application of Birkhoff's framework applied to the linearized form of \eqref{qP3}. When considering Birkhoff's fundamental solutions give solutions of \eqref{qP3}. In section \ref{sec:ultra}, by applying the ultradiscrete analog of this theory, we derive ultradiscrete fundamental solutions over $\Omega$ which map faithfully through the valuation to solutions of \eqref{udP3}. We also discuss the possibility of solutions in terms of Bessel functions and a method based purely on inequalities.

\section{Algebraic Tools}\label{sec:tools}

As mentioned above, we consider the lift of an equation to a field. The choice of valuation field depends on the application. One choice popular amongst tropical geometrists seems to be the field of algebraic functions with valuation defined to be the index of the pole or root at $0$ \cite{Introduction_Tropics}. In a previous paper, we used the lifting of a valuation field isomorphic to the inversible max-plus algebra (c.f.\cite{Ochiai_2005}), to derive some results regarding the solutions of linear systems over the tropical semiring $S$ \cite{Ormerod}. We choose to use the same construction. We shall give a standard definition of the max-plus semiring, $S$, then construct the valuation field $\Omega$.

Let $Q \subset \mathbb{R}$ be an additively closed and topologically closed subset of $\mathbb{R}$. We let $S$ be the set $Q \cup \{-\infty \}$ and equip $S$ with the natural ordering on $\mathbb{R}$. Here $-\infty$ is a minimal element. We equip $S$ with operations $\oplus$ and $\otimes$ which are defined by
\begin{subequations}\label{corr}
\begin{eqnarray}
a \oplus b = \max(a,b)\\
a \otimes b = a+ b
\end{eqnarray}
\end{subequations}
These operations are called tropical addition and tropical multiplication respectively. Here the $0$ element plays the role of the multiplicative identity while $-\infty$ plays the role of the additive identity. We may extend this definition to the set of matrices over $S$ in defining matrix operations $\oplus$ and $\otimes$. If we let $A = (a_{ij})$ and $B = (b_{ij})$, we may define matrix operations 
\begin{eqnarray*}
{}[A\otimes B]_{ij} &=& \max_k (a_{ik} + b_{kj}) \\
{}[A \oplus B]_{ij} &=& \max(a_{ij}, b_{ij}).
\end{eqnarray*}
with a scalar multiplication defined as
\[
{} [\lambda \otimes A]_{ij} = a_{ij} + \lambda
\]

The ultradiscretization of rational expressions and matrices can be seen as a homomorphism of the semiring $\mathbb{R}^+$ to the semiring $S$. Another way in which one may derive an expression over the max-plus semiring is to consider a non-archimedean valuation ring. 

\begin{definition}
A valuation ring is a ring $R$ with a valuation $\nu : R \to \mathbb{R}\cup \{-\infty \}$ such that 
\begin{enumerate}
\item{$\nu(x) = -\infty$ if and only if $x = 0$.}
\item{$\nu(xy) =  \nu(x) + \nu(y)$.}
\item{$\nu(x +y) \leq \nu(x)+\nu(y)$.}
\end{enumerate}
we call a valuation non archimedean if it has the property that
\begin{equation}\label{val}
\nu(x+y) \leq \max(\nu(x),\nu(y))
\end{equation}
\end{definition}

Suppose we consider a additively and multiplicatively closed subset $R_0 \subset R$ such that equality holds in equation \eqref{val} for all $x,y \in R_0$, then $\nu : R_0 \to S$ acts as a homomorphism of semirings. This allows full use of the ring $R$ to deduce properties of $R_0$ and then in turn $S$ via $\nu$. However, the equality of \eqref{val} may hold regardless of whether $x,y \in R_0$. This is the motivation for the current work. 

We now turn to the construction of the valuation field $\Omega$. Let $\Phi = \mathbb{Z}[G]$ be the set of formal $\mathbb{Z}$-linear combinations of a commutative group $G$. With typical elements $\chi,\phi \in \Phi$ represented by $\chi = \sum n_i (x_i)$ and $\phi = \sum m_i(y_i)$, the standard definition of the multiplication and addition in $\Phi$ is defined by
\begin{eqnarray*}
\chi + \phi &=& \sum_i n_i(x_i) + \sum_j m_j(y_j)\\
\chi \phi &=& \sum_{i,j} n_i m_j (x_i + y_j).
\end{eqnarray*}

We should note that the multiplicative identity element of the group is then denoted $1(0)$, the multiplicative identity of the group is denoted $0$ for the sum of no elements. We make the choice of group $G$ to be the group of reals $\mathbb{R}$ under addition. For our purposes, we consider $\Phi$ to be the set of $\mathbb{Z}$-linear combinations of $\mathbb{R}$ in which, for any element of $\Phi$, the real parts have an upper bound. We let $\Omega$ be the field of fractions of $\Phi$. We will typically identify any element of $\Omega$ as either a ratio of two elements of $\Phi$ or, if the element is contained within $\Phi$(\i.e. is of the form $\phi/1(0)$ for $\phi \in \Phi$), then we shall accordingly denote it as an element of $\Phi$. We define the valuation on $\Omega$ to be the mapping
\[
\nu \left(\frac{\sum n_i x_i}{\sum m_i y_i}\right) = \max(x_i) - \max(y_i).
\]
If $0 = \omega \in \Omega$ (i.e., an element in which the numerator is the sum of no elements), then $\nu(\omega)$ is the max of the empty set, which is $-\infty$ by definition. Let $\omega, \psi \in \Omega$ be given by
\begin{eqnarray*}
\omega = \frac{\sum n_{1,i}(x_{1,i})}{\sum n_{2,j}(x_{2,j})}\\
\psi = \frac{\sum m_{1,i}(y_{1,i})}{\sum m_{2,j}(y_{2,j})}
\end{eqnarray*}
then we have the identities
\begin{eqnarray*}
\nu(\omega \psi) &=& \nu\left( \frac{\sum n_{1,i}(x_{1,i})}{\sum n_{2,j}(x_{2,j})}\frac{\sum m_{1,i}(y_{1,i})}{\sum m_{2,j}(y_{2,j})}\right)\\
&=& \nu \left(\frac{ \sum_{ij} n_{1,i}m_{1,i}(x_{1,i} + y_{1,i}) }{\sum_{ij} n_{2,i}m_{2,i}(x_{2,i} + y_{2,i})}\right)\\
&=& \max_{ij}(x_{1,i} + y_{1,j}) - \max_{ij}(x_{2,i} + y_{2,j})\\
&=& \max(x_{1,i}) +\max(y_{1,i}) - \max(x_{2,i}) - \max(y_{2,i})\\
&=& \nu(\omega) + \nu(\psi)
\end{eqnarray*}
and 
\begin{eqnarray*}
\nu(\omega + \psi) &=& \nu\left( \frac{\sum n_{1,i}(x_{1,i})}{\sum n_{2,j}(x_{2,j})}+\frac{\sum m_{1,i}(y_{1,i})}{\sum m_{2,j}(y_{2,j})}\right)\\
&=& \nu\left( \frac{\sum n_{1,i}(x_{1,i})\sum m_{2,j}(y_{2,j})+\sum m_{1,i}(y_{1,i})\sum n_{2,j}(x_{2,j})}{\sum m_{2,j}(y_{2,j})\sum n_{2,j}(x_{2,j})}\right)\\
&=& \nu\left( \frac{\sum_{ij}n_{1,i} m_{2_j}(x_{1,i} + y_{2,j}) + \sum n_{2,i}m_{1,j}(x_{2,i} + y_{2,j})}{\sum_{ij} n_{2,i}m_{2,j}(x_{2,i}+y_{2,j})}\right)\\
&=& \max_{i,j}(x_{1,i} + y_{2,j}, x_{2,i} + y_{1,j}) - \max_{ij}(x_{2,i} + y_{2,j})\\
&=& \max_{i,j}(x_{1,i} + y_{2,j}- \max_{hk}(x_{2,h} + y_{2,k}), x_{2,i} + y_{1,j} - \max_{hk}(x_{2,h} + y_{2,k}))\\
&\leq& \max_{ij}(x_{1,i} + y_{2,j}- \max_{k}(x_{2,k} + y_{2,j}), x_{2,i} + y_{1,j} - \max_{k}(x_{2,i} + y_{2,k}))\\
&\leq& \max_{ij}(x_{1,i} - \max_k(x_{2,k}), y_{1,j} - \max(y_{2,k}))\\
&\leq& \max(\max_i(x_{1,i})-\max_k(x_{2,k}),\max( y_{1,j}) - \max_k(y_{2,k}))\\
&\leq& \max(\nu(\omega) , \nu(\psi)).
\end{eqnarray*}

These two identities are those required by a valuation, showing that this is indeed a non-archimedean valuation of $\Omega$. We have a lifting of any equation over $S$ to $\Omega$ via the mapping
\[
A \to 1(A) = 1(A)
\]
which we call the standard lift. It also has the property that the operations commute via the standard lift composed with $\nu$ (i.e. $\nu(1(A) + 1(B)) = A\oplus B$ and $\nu(1(A)1(B)) = A\otimes B$). This is the canonical choice of liftings, but it is worth noting that one tool of this theory is to consider liftings that are not standard, such as $A \to 1(A) + E$ where $\nu(1(A) + E) = A$. We also expect to lift any function, $F$ over $S$, to a function $\mathscr{F}$ over $\Omega$ where the only restriction is that $\nu(\mathscr{F}) = F$. We also note that if we restrict our attention to a special subsemiring given by
\[
\Omega_0 = \left\{ \frac{\sum n_i x_i}{\sum m_i y_i} | n_i,m_i > 0 \right\}
\]
then $\nu$ acts as a homomorphism of semirings. It is clear that any discrete dynamical system over $S$ has a corresponding lifted dynamical system over $\Omega_0$ in which anything that can be said over $\Omega_0$ can be said for $S$ through the valuation. 

\section{Derivation of $q$-hypergeometric solutions}\label{sec:hyper}
The continuous Painlev\'e equations possess hypergeometric solutions given by hypergeometric functions (and confluent hypergeometric functions). In the case of discrete equations, we require $q$-hypergeometric functions (and confluent $q$-hypergeometric functions) that come as generalizations of Hienes hypergeometric functions\cite{Kaji2}. Such equations then generalize to several variables giving an Askey-scheme of hypergeometric orthogonal polynomials and its q-analogs\cite{askey}. These are given by
\[
_r \phi_s\left(\begin{array}{c} a_1,\ldots, a_r \\ b_1,\ldots,b_s\end{array}| q : z \right) = \sum_{k = 0}^{\infty} \frac{(a_1,\dots, a_r ; q)_k}{(b_1,\dots, b_s ; q)_k} (-1)^{(1+r-s)k}q^{(1+r-s){k \choose 2}} \frac{z^k}{(q;q)_k}
\]
where 
\[
\begin{array}{c c c}
(a_1,\ldots, a_n;q)_k  = (a_1;q)\ldots(a_n;q) & \textrm{ and } & (a;q) = (1-a)(1-aq) \ldots (1-aq^{k-1})
\end{array}
\]
These functions have the property that they limit to hypergeometric function as $q \to 1$. We will predominantly be using the form where $r = s +1$, in which case the above form will simplify to the following
\[
_{s+1} \phi_s\left(\begin{array}{c} a_1,\ldots, a_{s+1} \\ b_1,\ldots,b_s\end{array}| q : z \right) = \sum_{k = 0}^{\infty} \frac{(a_1,\dots, a_{s+1} ; q)_k}{(b_1,\dots, b_s ; q)_k} \frac{z^k}{(q;q)_k}
\]
such a case is called well-posed. We begin by examining an equation in which the hypergeometric solutions are known. We start with the $q$-$\mathrm{P}_{III}$ given by
\begin{equation}\label{qP3}
\overline{w}\underline{w} = \frac{a_3 a_4(w + a_1 t)(w+a_2 t)}{(w+a_3)(w+a_4)}
\end{equation}
which was introduced in \cite{SC1}. This equation possesses an associated linear problem, or Lax pair, and is related to $q$-$\mathrm{P}_{VI}$ \cite{sakai}. The equation is known to have hypergeometric solutions of type $_2\phi_1$. If $a_2 = \frac{q a_1a_3}{a_4}$ then we find that \eqref{qP3} linearizes to give the following the Riccati equation
\begin{equation}\label{Rforward}
\overline{w} = -\frac{qta_1 a_3 + a_4 w}{a_3 + w}.
\end{equation}
We make the substitution $w = s x$ where $s = \sqrt{t}$ and $p = \sqrt{q}$ to derive the equation
\begin{equation}\label{Rbackward}
\overline{x} = -\frac{q s a_1 a_3 + a_4x}{a_3 + s x}.
\end{equation}

By using the Cole-Hopf transformation, $x = u/v$, we reduce this to the linear system
\begin{eqnarray*}
\begin{pmatrix} \overline{u} \\ \overline{v} \end{pmatrix} &=& \left(
\begin{pmatrix} a_4 & 0 \\
0 & a_3
\end{pmatrix} + 
\begin{pmatrix} 
0 & q a_1 a_3 \\
1 & 0
\end{pmatrix}s
\right)\begin{pmatrix} u \\ v \end{pmatrix}\\ \overline{\Psi} &=& (A_0 + A_1 s)\Psi.
\end{eqnarray*}

Explicit solutions in terms of $_2\phi_1$ functions, these can be obtained via an application of the work of LeCaine \cite{LeCaine_1943}. This shows the fundamental solution of $\Psi$ are 
\[
Y(x,t) = \Gamma_p\left(1+ \frac{t}{a_1a_3\sqrt{q}}\right)\begin{pmatrix}
\Phi_1(s) & \Phi_2(s) \\
\frac{\Phi_1(qs)}{\sqrt{a_1 a_4}} & \frac{\Phi_2(ps)}{a_3}\sqrt{\frac{a_4}{q a_1}} 
\end{pmatrix}\diag\left((-a_4/p)^{\log_p s},a_3^{\log_p s}\right)
\]
where the $\Phi$ functions are given by
\begin{eqnarray*}
\Phi_1(s) = _2\phi_1\left(\begin{array}{c} \frac{1}{\sqrt{a_1a_4}},a_3\sqrt{\frac{q a_1}{a_4}} \\ q \frac{a_3}{a_4}\end{array}| \sqrt{q}: s \right)\\
\Phi_2(s) = _2\phi_1\left(\begin{array}{c} \frac{1}{a_3}\sqrt{\frac{a_4}{q a_2}}, \sqrt{a_1a_4}\\\frac{a_4}{a_3} \end{array}| \sqrt{q}: s \right)
\end{eqnarray*}
and $\Gamma_q$ is the $q$-gamma function in \cite{askey}. From this, we may express $w = u/v$ explicitly in terms of $\Phi_1$ and $\Phi_2$. The other approach is to show that under certain conditions, by substituting a different form to \eqref{Riccati}, one can find solutions that can be expressed in terms of $q$-Bessel functions. The details of this method are given in \cite{Kaji2}.

\section{Ultradiscrete hypergeometric solutions}\label{sec:ultra}

The question we wish to address, is whether something can be said for the correspondence with these explicit solutions and the ultradiscrete version of the same equation. Ideally, one would expect that if one has a solution of \eqref{qP3}, then the ultradiscretization of such a solution of yields a solution of \eqref{udP3}. Yet since the evolution of the solution given by \eqref{Rforward} and \eqref{Rbackward} involves a negative sign, it is clear, we must leave the subtraction free framework. If we lift the ultradiscretization of \eqref{qP3}, and apply the valuation, one could expect cases in which the valuation yields new solutions which will be ultradiscrete analogs of the hypergeometric functions.

The ultradiscretization method applied to \eqref{qP3} gives us the discrete dynamical system
\begin{eqnarray*}
\overline{W} +\underline{W} &=& \max(W, A_1 + T)+ \max(W,A_2 + T) \\
&& - \max(W,A_3) - \max(W,A_4) \nonumber.
\end{eqnarray*}

The equation is known to possess rational solutions such as those obtained via orbits of the group of B\"acklund transformations \cite{ultimate} applied to the solution $W = T/2$ for $A_1 = A_2 = A_3 = A_4 =0$. It is also known to admit an Affine Weyl group representation as a group of B\"acklund transformations. By following such a procedure, one obtains solutions that are simply the rational solutions that one may obtain through the ultradiscretization of known rational solutions of \eqref{qP3}.

We lift equation \eqref{udP3} to $\Omega$ via the standard lift. The lifting gives the equation
\begin{equation}\label{ludP3}
\overline{\mathscr{W}}\underline{\mathscr{W}} = \frac{1(A_3)1(A_4)(\mathscr{W} + 1(A_1+T))(\mathscr{W}+ 1(A_2+T))}{(\mathscr{W} + 1(A_3))(\mathscr{W}+1(A_4))}
\end{equation}
If $\mathscr{W}$ is an element of $\Omega_0$, then we have that the valuation $\nu$ brings such an equation to \eqref{udP3}. We note that any rational expression over $\Omega_0$ is the manifestation of a subtraction free expression over the reals, making it apparent that the set of rational solutions can be expressed as the mappings of solutions of the lifted equation that exist in $\Omega_0$. For general $\mathscr{W}$ however, we have a set of inequalities obtained by breaking up the parts. Given $\nu(\mathscr{W}) = W$ ,we have the following list of inequalities
\begin{subequations}\label{ineqexample}
\begin{eqnarray}
\nu(\mathscr{W}+ 1(A_1+T)) &\leq& \max(W,A_1+T)\\
\nu(\mathscr{W}+ 1(A_2+T)) &\leq& \max(W,A_2+T)\\
\nu(\mathscr{W}+ 1(A_3)) &\leq& \max(W,A_3)\\
\nu(\mathscr{W}+ 1(A_4)) &\leq& \max(W,A_4).
\end{eqnarray}
\end{subequations}

Although this does not provide upper or lower bounds for $\overline{W}$ or $\underline{W}$, imposing equality is a relaxation of the subtraction free framework. Note that we may relax this condition further by imposing 
\begin{eqnarray}\label{eq:less}
(\nu(\mathscr{W}+ 1(A_i+T))- \max(W_i,A_1+T))  + \\ 
(\nu(\mathscr{W}+ 1(A_2+T))- \max(W_i,A_2+T)) - \nonumber \\
(\nu(\mathscr{W}+ 1(A_3)) - \max(W_i,A_3)) - \nonumber \\
(\nu(\mathscr{W}+ 1(A_4)) - \max(W_i,A_4)) = 0 \nonumber 
\end{eqnarray}
which we refer to as an equality of certain differences. This lifting is also rather artificial way of regaining singularities in ultradiscrete systems, where any lift of an ultradiscrete system can possess singularities in $-\Omega_0$. In the tropical sense however, we consider the images of the singularities under $\nu$ to be tropical singularities, which is also the set of points where a max-plus function is not linear. This also coincides with the singularity in terms of ultradiscrete singularity confinement \cite{Joshi_2005}. 

If it is possible to show the equality or equality of the difference of arguments in \eqref{ineqexample} or in the less restrictive \eqref{eq:less}, then we have an expression for some solution $W$ of \eqref{udP3}. The crux of this method is to consider the possibility of considering this equality outside of $\Omega_0$. That is to say that it is possible satisfy this requirement over for expressions in $\Omega$ but not necessarily in $\Omega_0$.

As an example, let us consider the linearization of \eqref{ludP3}. We should be able to derive solutions in a similar manner to the work above for \eqref{qP3}. We substitute the following lifted ultradiscretized Riccati equation over $\Omega$
\begin{equation}\label{Riccati}
\overline{\mathscr{W}} = \frac{1(\alpha) + 1(\beta)\mathscr{W}}{1(\gamma) + 1(\delta)\mathscr{W}}
\end{equation}
We find conditions for \eqref{Riccati} to describe the evolution of \eqref{ludP3}. These conditions coincide with the ultradiscretized conditions for the linearization of \eqref{qP3} for similar reasons. These are that $A_2 = Q + A_1 +A_3-A_4$, giving the linear system over $\Omega$
\begin{subequations}\label{evol}
\begin{eqnarray}
\label{fevolR}\overline{\mathscr{W}} = \frac{-1(A_1 + A_3 +T+ Q) - 1(A_4)\mathscr{W}}{1(A_3) + \mathscr{W}}\\
\label{bevolR}\underline{\mathscr{W}} = \frac{-1(A_1 + A_3 +T) - 1(A_3)\mathscr{W}}{1(A_4) + \mathscr{W}}.
\end{eqnarray}
\end{subequations}

We derive inequalities for \eqref{fevolR}, these are
\begin{subequations}\label{fineq}
\begin{eqnarray}
\nu (-1(A_1 + A_3 +T+ Q) - 1(A_4)\mathscr{W}) &\leq& A_4 + \max(A_2 + T, W)\\
\nu (1(A_3) + \mathscr{W}) &\leq& \max(A_3,W)
\end{eqnarray}
\end{subequations}
and for \eqref{bevolR} we have
\begin{subequations}\label{bineq}
\begin{eqnarray}
\nu(-1(A_1 + A_3 +T) - 1(A_3)\mathscr{W}) &\leq& A_3 + \max(A_1 + T, W)\\
\nu(1(A_4) + \mathscr{W}) &\leq& \max(A_4,W)
\end{eqnarray}
\end{subequations}
where in each case, if equality, or equality of certain differences hold for \eqref{fineq} and \eqref{bineq}, then $\nu(W)$ is a solution of \eqref{udP3}. Given appropriate conditions, it is possible to make it so this equality hold for all time. If equality does hold, then the positive and negative evolution is be given simply by two ultradiscrete Riccati type equations for each direction
\begin{subequations}\label{Sevol}
\begin{eqnarray}
\label{Sevol:fore}\overline{W} = \max(A_1 + A_3 + T + Q, A_4 + W) - \max(A_3,W)\\
\label{Sevol:back}\underline{W} = \max(A_1 + A_3 + T,A_3 + W) - \max(A_4,W).
\end{eqnarray}
\end{subequations}

\begin{figure}[!ht]
\subfigure[\label{hyper1}Hypergeometric solution for parameters $A_1 = -2, A_2 = -1, A_3 =0, A_4 =0, Q = 1$ and $W_0(0) = 0$.]{
\includegraphics[width = 6cm]{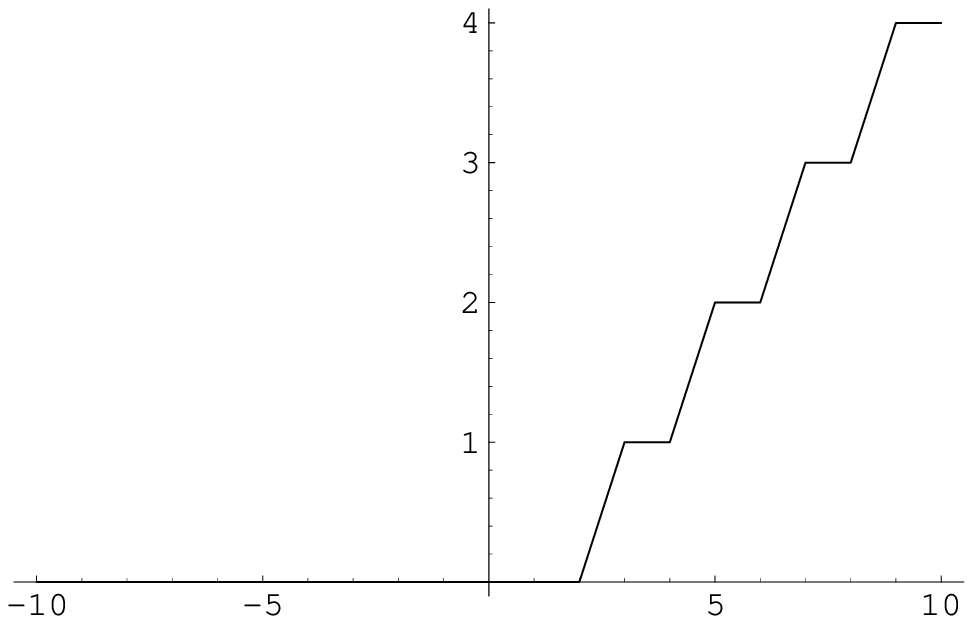}}
\subfigure[Hypergeometric solution for parameters $A_1 = 0, A_2 = 2, A_3 =1, A_4 =0, Q = 1$ and $W_0(0) = 1$.]{
\includegraphics[width = 6cm]{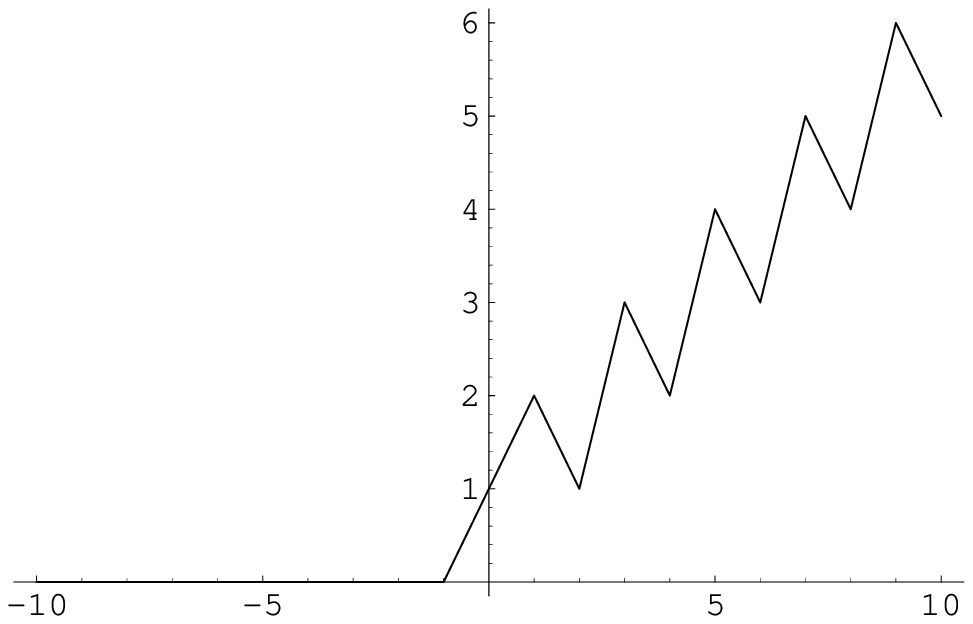}}
\caption{\label{example}2 Hypergeometric solutions to \eqref{udP3}.}
\end{figure}

Due to the way in which the minus sign in \eqref{Rforward} and \eqref{Rbackward} appears, it is clear that any resultant solution is not any solution that can come from the ultradiscretization of a solution of \eqref{qP3} but rather a hypergeometric solution of \eqref{udP3} which is unique to the semiring setting. Figure \ref{example} shows two examples of evolutions in which the functions defined by \eqref{Sevol} and \eqref{udP3} coincide.

One particular family of systems is defined by the conditions 
\begin{eqnarray*} 
A_1 = -(r+1)Q &  A_2 = Q + A_1 +A_3-A_4 = -rQ\\
A_3 = 0 & A_4 = 0
\end{eqnarray*}
where $W(0) = 0$ and $r >1$. This particular example we will call a hypergeometric solution of \eqref{udP3}. From here on, we shall be considering this example in detail from various standpoints.

For $T < 0$, and $W = 0$, we have an evolution giving $0$ for all time, which is consistent with the evolution of \eqref{ludP3}, however for $T > 0$, since $A_1 \leq - 2Q$, using \eqref{Sevol}, $W(T) < A_2 + T$ implies that $\overline{W} \leq A_2 + T + Q$ and similarly we obtain the same for $A_1$. We also have that for $T > A_2 > A_1$, then $W(T) > 0 = A_3,A_4$ giving us all inequalities required to derive the precise evolution of \eqref{udP3}. From these inequalities, we obtain the precise evolution 
\[
\overline{W} + \underline{W} = A_3 + A_4 + A_1 + A_2 + 2T - 2W
\]
for $T > A_2 > A_1$. We may also derive the exact same set of inequalities using \eqref{udP3}; hence we arrive at the same equation governing the evolution of \eqref{Sevol} and \eqref{udP3}. To derive the precise form of resultant solution, we may either solve this as a difference equation in terms of $\frac{nQ}{2}$, and $(-1)^n$, or alternatively we may use the theory of linear difference equations over the semiring. From the perspective of semirings, it is then clear that the Riccati forms given by \eqref{Sevol} define the evolution for \eqref{udP3}.

Over $\Omega$ however, given a solution in which the evolution governed by \eqref{udP3} and \eqref{Sevol} coincide, we consider the lifted equation, \eqref{evol}, as also having a solution defined over $\Omega$ which may be considered an analog of the hypergeometric function over a field. We shall be able to derive explicit expressions for our new ultradiscrete hypergeometric functions through this correspondence and $\nu$. Since the expression for the evolution is given by \eqref{evol}, we can use a Cole-Hopf transformation over $\Omega$. By substituting $\mathscr{W} = U/V$, we derive the linear system for specific examples that satisfy the prerequisites for the existence of the hypergeometric solution. From \eqref{evol} and under these assumptions, we have the linear system
\[
\begin{pmatrix}\overline{U} \\ \overline{V}\end{pmatrix} = \begin{pmatrix} -1(0) & -1(T-rQ) \\ 1(0) & 1(0) \end{pmatrix} \begin{pmatrix} U \\ V\end{pmatrix} = Y(1(T)1(Q)) = A(1(T))Y(1(T))
\]

This is a linear system in which the framework of \cite{Ormerod} applies. We may solve this explicitly by substituting and expression of the form
\[
Y(1(T)) = (I + Y_1 1(T) + Y_2 1(2T) + \ldots )\diag(-1(0)^{\frac{T}{Q}}, 1(0))
\]
from which we derive the recursion relation 
\[
Y_{k+1} = (\diag(1(Q)^{k+1}, 1(Q)^{k+1}) - A_0) A_1 Y_{k}).
\] 
Solving this recursion relation, we arrive at the expansion
\begin{equation}\label{coeff}
Y_k = \begin{pmatrix}
0 & \frac{1(Q)^{-kr} (1(Q)^k - 1(0))}{(1(Q):1(Q))_{k+1}(-1(Q);1(Q))} \\
0 & \frac{1(Q)^{-kr}}{(1(Q):1(Q))_k(-1(Q);1(Q))}
\end{pmatrix}
\end{equation}
where we have the Pochammer symbol over $\Omega$ is 
\[
(A;B) = (1 - A)(1 - A B^2) \ldots (1 - A B^{k-1}).
\]

If we follow through the hypergeometric representation of these objects over $\Omega$, we may express the solution as
\[
Y(t) = \begin{pmatrix}
1(0) & _2\phi_1 \left( \begin{array}{c}0, 0 \\ -1(0) \end{array}| 1(Q); 1(T-rQ) \right)-1(0) \\
0 & _2 \phi_1 \left( \begin{array}{c}0, 0 \\ -1(Q) \end{array}| 1(Q); 1(T-rQ) \right) \\
\end{pmatrix}.
\]

By stating explicitly the initial condition of $(1(0),1(0))$ the solution is simply the ratio of these hypergeometric functions. We calculate the valuation of this solution from \eqref{coeff}, and the valuation of hypergeometric solution results in the following formula for the hypergeometric
over $S$
\begin{equation}\label{solution}
\nu\left( \frac{_2\phi_1 \left( \begin{array}{c}0, 0 \\ -1(0) \end{array}| 1(Q); 1(T-rQ) \right)}{_2 \phi_1 \left( \begin{array}{c}0, 0 \\ -1(Q) \end{array}| 1(Q); 1(T-rQ) \right)}\right) = \begin{array}{c} 
\max_{k \in \mathbb{N}} (kQ - k r Q-(k+1)kQ+kT)- \\  
\max_{k \in \mathbb{N}} (-k r Q-(k+1)kQ+kT)\end{array}
\end{equation}
which for all $r > 1$, we shall show this coincides precisely with the calculation of the solution given by calculating the evolution via \eqref{udP3} given the appropriate initial conditions. To see this, we simplify this expression by letting $T = nQ$ and hence the solution is expressed as
\begin{equation}\label{eq:simpw}
W = Q \max_k (k(n-r-k)) - Q \max_k (k(n-r-k-1)) 
\end{equation}
where $\overline{W}(n)$ is now $W(n+1)$. We see that $W \geq 0$ for all $n$, hence we simplify \eqref{Sevol:fore} to
\[
\overline{W} = \max(Q(n-r) - W,0)
\]
which we substitute \eqref{eq:simpw} 
\begin{eqnarray*}
Q\max_k(k(n+1-r-k)) &= &Q\max\left((n-r) + \max_k (k(n-r-k-1))\right. \\
- Q \max_k (k(n-r-k)) && -\left. \max_k (k(n-r-k)),0\right)\\
&=& Q \max_k\left( (k+1)n - (k+1)r -k^2 - k, k(n-r-k) \right)\\
&&  -Q\max_k (k(n-r-k)).
\end{eqnarray*}
By shifting the $k$ component in the first max expression, we write
\begin{eqnarray*}
\overline{W} &=& Q \max_{k \geq 1} \left(kn-kr-(k-1)^2 -(k+1) ,kn - kr -k^2-k ,0)\right)\\
&&  -Q\max_k (k(n-r-k))\\
&=& Q \max_{k \geq 1} \left( kn - kr -k^2 +k, kn - kr -k^2-1 ,0 \right)\\
&&  -Q\max_k (k(n-r-k)).
\end{eqnarray*}

We may simplify the right hand side way way of noticing that the first elements in the max always dominates for $k \geq 1$.
This gives the equality 
\begin{eqnarray*}
\overline{W} &=& Q \max_{k \geq 1} \left(kn-kr-k^2+k,0)\right)\\
&&  -Q\max_k (k(n-r-k))\\
&=& \max_k \left( k(n-r-k+1)\right) - \max_k (k(n-r-k))
\end{eqnarray*}
which shows the \eqref{eq:simpw} does satisfy \eqref{Sevol:fore}. To see that this equation satisfies \eqref{Sevol:back}, 
since the forward equation holds, we may de-evolve the forward evolution equation, thus we write
\[
W = \max((n-r-1)Q - \underline{W},0) 
\]
in which if $W > 0$ then $W = (n-r-1)Q - \underline{W}$. We also know that $W \geq \underline{W} \geq 0$ hence for $W =0$, $\underline{W} = 0$ thus we have 
\[
\underline{W} = \max((n-r-1)Q - W, 0)
\]
for all $n$. This which coincides with \eqref{Sevol:back} for our choice of $W$ and our parameters hence \eqref{eq:simpw} solves \eqref{Sevol}. This then immediately implies \eqref{eq:simpw} solves u-$\mathrm{P}_{\mathrm{III}}$.

This shows that not only does such a hypergeometric solution have an evolution defined by some Riccati equation, but it also can be derived as a valuation of a hypergeometric equation in a higher space. An alternative approach to that of the linear systems approach is to consider the above family in terms of $q$-Bessel functions similar to that of \cite{Kaji2}. Instead of the Cole-Hopf transformation, we may consider a different substitution into \eqref{fevolR} given by
\[
\mathscr{W} = \frac{\overline{J}}{J}-1(0)
\]
In which then under the condition $A_3 = 1(0),A_4=-1(0)$ and $A_1 = (1(-Q)-1(0))^2$, then must satisfy the equation
\[
\overline{J}-J(1(0)+ 1(0))+ \underline{J}(1(0)+1(T-rQ)A_1)=0
\]
which is then a lifted ultradiscretized form of the Bessel equation for $J_\nu(S)$ where $S^2 = T$. This is a special case where $\nu =0$, the resultant solution of this equation is the Bessel over $\Omega$. This function simplifies in the case $\nu = 0$ to the equation
\[
J_0(T;Q) =  = _2 \phi_1 \left( \begin{array}{c}0, 0 \\ 1(Q) \end{array}| 1(Q); 1(T-rQ) \right)
\]
substituting back into $\mathscr{W}$, and applying the valuation with appropriate initial conditions, we recover the same formula for the solution as found by our previous method with linear systems in \eqref{solution}.

Although these two methods yield the same expression for the solution, over $\Omega$ they are solutions to two different equations, both whose mapping via $\nu$ give \eqref{Sevol}. Furthermore, all three methods give the same derivation of the behavior seen in figure \ref{hyper1}. If one can tell a priori via some inductive or deductive process whether the equality of \eqref{val}, then it would allow us to see more clearly what other solutions to ultradiscrete equations do not come as ultradiscretized subtraction free solutions. We propose that if this could be done, perhaps it would be fair to say that this process yields a new ultradiscretization method different from \cite{Takahashi_1997} and \cite{sinh}. It also puts ultradiscretization in a sense, in terms of non-archimedean valuations of a field. 

\section{Conclusion}
Despite the existence of hypergeometric type solutions, the conditions in which the valuation yields equality of both sides is unclear when considering a solution holding for all time, making it difficult to assert when this method will give rise to new solutions. Further investigation into these conditions may give rise to other types of solutions, including the precise conditions for when a ultradiscrete equation has what we consider a hypergeometric solution. In particular, it would be interesting to see whether solutions may be parameterized in terms of a general ultradiscrete Bessel function.

This method opens a door in allowing us to consider studies in ultradiscrete equations outside the ultradiscretization of any subtraction free method applied to $q$-difference equations. In particular, it allows us to consider solutions of various ultradiscrete equations in which the solution does not come as an ultradiscretized subtraction free solution of the $q$-difference equation it was derived from.

\section{Acknowledgement}
The author would like to acknowledge the contributions of Professor Kenji Kajiwara, Doctor Chris Field and Professor Yasuhiko Yamada for their helpful points and insights. This research was supported in part by the Australian Research Council grant \#DP0559019.

\end{document}